\documentclass{ws-ijmpe}
\usepackage{graphicx}
\begin{document}

\markboth{Ryo Takahashi}{Dark energy and neutrino model in SUSY\\
-- Remarks on active and sterile neutrinos mixing --}

\catchline{}{}{}{}{}
%%%%%%%%%%%%%%%%%%%%%%%%%%%%%%%%%%%%%%%%%%%%%%%%%%%%%%%%%%%%%%%%%%%%

\title{Dark energy and neutrino model in SUSY\\
-- Remarks on active and sterile neutrinos mixing --}

\author{\footnotesize Ryo Takahashi\footnote{talked at the
International Workshop on Neutrino Masses and Mixings, University of
Shizuoka, Shizuoka, Japan, December 17-19, 2006}}

\address{Graduate School of  Science and Technology,
 Niigata University, 950-2181 Niigata, Japan\\
takahasi@muse.sc.niigata-u.ac.jp}

\author{Morimitsu Tanimoto}

\address{Department of Physics,
 Niigata University,  950-2181 Niigata, Japan\\
tanimoto@muse.sc.niigata-u.ac.jp}

\maketitle

%\begin{history}
%\received{(received date)}
%\revised{(revised date)}
%\accepted{(Day Month Year)}
%\comby{(xxxxxxxxxx)}
%\end{history}

\begin{abstract}
We consider a Mass Varying Neutrinos (MaVaNs) model in supersymmetric
theory. The model includes effects of supersymmetry breaking
transmitted by the gravitational interaction from the hidden sector,
in which supersymmetry was broken, to the dark energy sector. Then
evolutions of the neutrino mass and the equation of state parameter of
the dark energy are presented in the model. It is remarked that only
the mass of a sterile neutrino is variable in the case of the
vanishing mixing between the left-handed and a sterile neutrino on
cosmological time scale. The finite mixing makes the mass of the
left-handed neutrino variable.  
\end{abstract}

\section{Introduction}

Cosmological observations have provided the strong evidence that the
Universe is flat and its energy density is dominated by the dark
energy component whose negative pressure causes the cosmic expansion
to accelerate.\cite{Riess} In order to clarify the origin of the dark
energy, one has tried to understand the connection of the dark energy
with particle physics.

In the Mass Varying Neutrinos (MaVaNs) scenario proposed by Fardon,
Nelson and Weiner, relic neutrinos could form a negative pressure
fluid and cause the present cosmic acceleration.\cite{Weiner} In the
model, an unknown scalar field, which is called ``acceleron'', is
introduced and neutrinos are assumed to interact through a new scalar
force. The acceleron sits at the instantaneous minimum of its
potential, and the cosmic expansion only modulates this minimum
through changes in the number density of neutrinos. Therefore, the
neutrino mass is given by the acceleron, in other words, it depends on
the number density of neutrinos and changes with the expansion of the
Universe. The equation of state parameter $w$ and the energy density
of the dark energy also evolve with the neutrino mass. Those
evolutions depend on the form of a scalar potential and the relation
between the acceleron and the neutrino mass strongly. Some examples of
the potential have been considered.\cite{Peccei}

The idea of the variable neutrino mass was considered at first in a
model of neutrino dark matter and was discussed for neutrino
clouds.\cite{Yanagida} Interacting dark energy scalar with neutrinos
was considered in the model of a sterile neutrino.\cite{Hung} The
coupling to the left-handed neutrino and its implication on the
neutrino mass limit from baryogenesis was discussed.\cite{Wang} In
the context of the MaVaNs scenario, there have been a lot of works.
\cite{Kaplan,our,Honda,hybrid,stability,Speed}

In this talk, we present a MaVaNs model including the supersymmetry
breaking effect mediated by the gravity. Then we show evolutions of
the neutrino mass and the equation of state parameter in the model.

\section{MaVaNs Model in Supersymmetric Theory}

We discuss the Mass Varying Neutrinos scenario in supersymmetric
theory and present a model.

We assume a chiral superfield $A$ in dark sector. $A$ is assumed to be
a singlet under the gauge group of the standard model. It is difficult
to construct a viable MaVaNs model without fine-tunings in some
parameters when one assumes one chiral superfield in dark sector,
which couples to only the left-handed lepton doublet superfield.
\cite{our} Therefore, we assume that the superfield $A$ couples to
both the left-handed lepton doublet superfield $L$ and the
right-handed neutrino superfield $R$. For simplicity, we consider the
MaVaNs scenario in one generation of neutrinos.\footnote{Three
generations model of this scenario has presented in non supersymmetric 
theory.\cite{Honda}}

In such framework, we suppose the following superpotential,
 \begin{eqnarray}
  W=\frac{\lambda}{6}A^3+\frac{M_A}{2}AA+m_DLA+M_DLR
    +\frac{M_R}{2}RR,\label{W}
 \end{eqnarray}
where $\lambda$ is a coupling constant of $\mathcal{O}(1)$ and $M_A$, 
$M_D$, $M_R$ and $m_D$ are mass parameters.\footnote{Other
 supersymmetric model so called ``hybrid model'' has been
 proposed.\cite{hybrid}} The scalar and the spinor component of $A$
 are represented by $\phi$ and $\psi$, respectively. The scalar
 component corresponds to the acceleron which cause the present cosmic
 acceleration. The spinor component is a sterile neutrino. The third
 term of the right-hand side in Eq. (\ref{W}) is derived from the
 Yukawa coupling such as $yLAH$ with $y<H>=m_D$, where $H$ is the
 Higgs doublet. 

In the MaVaNs scenario, the dark energy is assumed to be composed of
the neutrinos and the scalar potential for the acceleron. Therefore,
the energy density of the dark energy is given as
 \begin{equation}
  \rho _{\mbox{{\scriptsize DE}}}=\rho _\nu +V(\phi ).
 \end{equation}
Since only the acceleron potential contributes to the dark energy, we
assume the vanishing vacuum expectation values of sleptons, and thus
we find the following effective scalar potential,
 \begin{equation}
  V(\phi )=\frac{\lambda^2}{4}|\phi |^4+M_A^2|\phi |^2
           +m_D^2|\phi |^2.
 \end{equation}
We can write down a lagrangian density from Eq. (\ref{W}),
 \begin{eqnarray}
  \mathcal{L}=\lambda\phi\psi\psi +M_A\psi\psi+m_D\nu_L\psi
              +M_D\nu_L\nu _R+M_R\nu _R\nu _R+h.c.. 
  \label{lag}
 \end{eqnarray}
It is noticed that the lepton number conservation in the dark sector
is violated because this lagrangian includes both $M_A\psi\psi$ and
$m_D\nu_L\psi$. After integrating out the right-handed neutrino, the
effective neutrino mass matrix is given by
 \begin{eqnarray}
  \mathcal{M}\simeq
   \left(
   \begin{array}{cc}
    c   & m_D \\
    m_D & M_A+\lambda\phi
   \end{array}
  \right),
  \label{MM}
 \end{eqnarray}
in the basis of $(\nu_L,\psi )$, where $c\equiv -M_D^2/M_R$ and we
assume $\lambda\phi\ll M_D\ll M_R$. The first term of the $(1,1)$
element of this matrix corresponds to the usual term given by the
seesaw mechanism in the absence of the acceleron. We obtain masses of
the left-handed and a sterile neutrino as follows,
 \begin{eqnarray}
  m_{\nu _L}&=&\frac{c+M_A+\lambda<\phi>}{2}
               +\frac{\sqrt{[c-(M_A+\lambda<\phi>)]^2+4m_D^2}}{2},\\
  m_\psi&=&\frac{c+M_A+\lambda<\phi>}{2}
           -\frac{\sqrt{[c-(M_A+\lambda<\phi>)]^2+4m_D^2}}{2}.
 \end{eqnarray}
 It is remarked that
only the mass of a sterile neutrino is variable in the case of the
vanishing mixing ($m_D=0$) between the left-handed and a sterile
neutrino on cosmological time scale. The finite mixing ($m_D\neq 0$)
makes the mass of the left-handed neutrino variable. We will consider
these two cases of $m_D=0$ and $m_D\neq 0$ later.

In the MaVaNs scenario, there are two constraints on the scalar
potential. The first one comes from cosmological observations. It is
that the magnitude of the present dark energy density is about
$0.74\rho _c$. $\rho _c$ is the critical density. Thus, the first
constraint turns to
 \begin{eqnarray}
  V(\phi ^0)=0.74\rho _c-\rho _\nu ^0,
  \label{V}
 \end{eqnarray}
where ``$0$'' means the present value.

The second one is the stationary condition:
 \begin{eqnarray}
  \frac{\partial\rho_{\mbox{{\scriptsize DE}}}}{\partial\phi}
  =\frac{\partial\rho_\nu}{\partial\phi}
   +\frac{\partial V(\phi)}{\partial\phi}=0.
  \label{7}
 \end{eqnarray}
In this scenario, the neutrino mass is represented by a function of
 the acceleron; $m_\nu=f(\phi)$. Since the energy density of the 
neutrino varies on cosmological times scale, the vacuum 
expectation value of the acceleron also varies. This property makes 
the neutrino mass variable. If $\partial m_\nu/\partial\phi\neq0$, 
Eq. (\ref{7}) is equivalent to
 \begin{eqnarray}
  \frac{\partial\rho _{\mbox{{\scriptsize DE}}}}{\partial m_\nu}
  =\frac{\partial\rho _\nu}{\partial m_\nu}
   +\frac{\partial V(\phi (m_\nu))}{\partial m_\nu}=0.
  \label{stationary}
 \end{eqnarray}
Eq. (\ref{stationary}) is rewritten by using the cosmic temperature $T$:
 \begin{eqnarray}
  \frac{\partial V(\phi )}{\partial m_\nu}
  =-T^3\frac{\partial F(\xi )}{\partial\xi},
  \label{stationary1}
 \end{eqnarray}
where $\xi\equiv m_\nu /T$, $\rho _\nu =T^4F(\xi )$ and
 \begin{eqnarray}
  F(\xi )\equiv\frac{1}{\pi ^2}\int _0^\infty
               \frac{dyy^2\sqrt{y^2+\xi ^2}}{e^y+1}.
 \end{eqnarray}
We can get the time evolution of the neutrino mass from
Eq. (\ref{stationary1}). Since the stationary condition should be
always satisfied in the evolution of the Universe, this one at the
present epoch is the second constraint on the scalar potential:
 \begin{eqnarray}
  \left.\frac{\partial V(\phi )}{\partial m_\nu}\right|
  _{m_\nu =m_\nu^0}
  =\left.-T^3\frac{\partial F(\xi )}{\partial\xi}\right|
   _{m_\nu =m_\nu^0,T=T_0}.
  \label{stationary2}
 \end{eqnarray} 
In addition to two constraints for the potential, we also have two
relations between the vacuum expectation value of the acceleron and the neutrino masses at the present
epoch:
 \begin{eqnarray}
  m_{\nu _L}^0&=&\frac{c+M_A+\lambda<\phi>^0}{2}
                 +\frac{\sqrt{[c-(M_A+\lambda<\phi>^0)]^2+4m_D^2}}
                       {2},\\
  m_\psi ^0&=&\frac{c+M_A+\lambda<\phi>^0}{2}
              -\frac{\sqrt{[c-(M_A+\lambda<\phi>^0)]^2+4m_D^2}}{2}.
 \end{eqnarray}

Next, let us consider the dynamics of the acceleron field. In order
that the acceleron does not vary significantly on distance of
inter-neutrino spacing, the acceleron mass at the present epoch must
be less than $\mathcal{O}(10^{-4}\mbox{eV})$ \cite{Weiner}. Here and
below, we fix the present acceleron mass as
 \begin{eqnarray}
  m_\phi ^0=10^{-4}\mbox{ eV}.
  \label{amass}
 \end{eqnarray}
Once we adjust parameters which satisfy five equations (\ref{V}) and
(\ref{stationary2})$\sim$(\ref{amass}), we can have evolutions of the
neutrino masses by using the Eq. (\ref{stationary1}).  

The dark energy is characterized by the evolution of the equation of
state parameter $w$. The equation of state is derived from the energy
conservation law and the stationary condition Eq. (\ref{stationary1}):
 \begin{eqnarray}
  w+1=\frac{[4-h(\xi )]\rho _\nu}{3\rho _{\mbox{{\scriptsize DE}}}},
 \end{eqnarray}
where
 \begin{eqnarray}
  h(\xi )\equiv\frac{\xi\frac{\partial F(\xi )}{\partial\xi}}{F(\xi )}.
 \end{eqnarray}
It seems that $w$ in this scenario depend on the neutrino mass and the
cosmic temperature. This means that $w$ varies with the evolution of
the Universe unlike the cosmological constant.

In the last of this section, we comment on the hydrodynamical
stability of the dark energy in the MaVaNs scenario. The speed of
sound squared in the neutrino-acceleron fluid is given by
 \begin{eqnarray}
  c_s^2=\frac{\dot{p}_{\mbox{{\scriptsize DE}}}}
             {\dot{\rho}_{\mbox{{\scriptsize DE}}}}
       =\frac{\dot{w}\rho _{\mbox{{\scriptsize DE}}}
              +w\dot{\rho}_{\mbox{{\scriptsize DE}}}}
             {\dot{\rho}_{\mbox{{\scriptsize DE}}}},
 \end{eqnarray}
where $p_{\mbox{{\scriptsize DE}}}$ is the pressure of the dark
energy. Recently, it was argued that when neutrinos are
non-relativistic, this speed of sound squared becomes negative in this
scenario.\cite{stability} The emergence of an imaginary speed of
sound means that the MaVaNs scenario with non-relativistic neutrinos
is unstable, and thus the fluid in this scenario cannot acts as the
dark energy. However, finite temperature effects provide a positive
contribution to the speed of sound squared and avoid this instability.
\cite{Speed} Then, a model should satisfy the following condition,
 \begin{eqnarray}
  \frac{\partial m_\nu}{\partial z}
  \left(1-\frac{5aT^2}{3m_\nu ^2}\right)
  +\frac{25aT_0^2(z+1)}{3m_\nu}>0,
  \label{constraint}
 \end{eqnarray}
where $z$ is the redshift parameter, $z\equiv (T/T_0)-1$, and 
 \begin{eqnarray}
  a\equiv\frac{\int _0^\infty\frac{dyy^4}{e^y+1}}
              {2\int _0^\infty\frac{dyy^2}{e^y+1}}\simeq 6.47.
 \end{eqnarray}
The first and the second term of left hand side in Eq. 
(\ref{constraint}) are negative and positive contributions to the
speed of sound squared, respectively. We find that a model which leads
to small $\partial m_\nu/\partial z$ is favored. A model with a small
power-law scalar potential; $V(\phi)=\Lambda^4(\phi/\phi^0)^k$,
$k\ll1$, and a constant dominant neutrino mass; $m_\nu=C+f(\phi)$,
$f(\phi)\ll C$, leads to small $\partial m_\nu/\partial
z$.\footnote{A model with the masses of the left-handed neutrinos
given by the see-saw mechanism is unstable even if it has a small
power-law scalar potential.\cite{Spitzer}} Actually, some models have
been presented.\cite{Honda}

\begin{figure}[t]
\begin{center}
\includegraphics[width=0.5\linewidth]{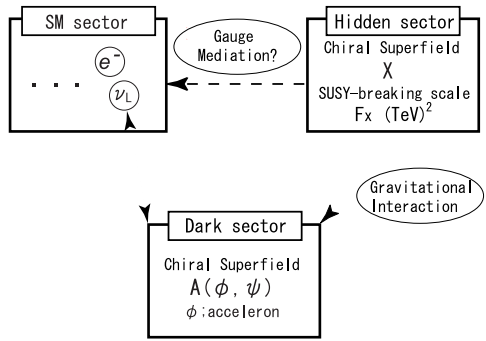}
\end{center}
\caption{The illustration of interactions among three sectors.
The dark sector couples to the left-handed neutrino through a new 
scalar force in the MaVaNs scenario. The dark sector is also assumed
to be related with the hidden sector only through the gravity.}
\label{fig:0}
\end{figure}

\section{Effect of supersymmetry breaking}
Let us consider effect of supersymmetry breaking in the dark
sector. We assume a superfield $X$, which breaks supersymmetry, in the 
hidden sector, and the chiral superfield $A$ in the dark sector is
assumed to interact with the hidden sector only through the
gravity. This framework is shown graphically in Fig. \ref{fig:0}. Once
supersymmetry is broken at TeV scale, its effect is transmitted to the
dark sector through the following operators:
 \begin{eqnarray}
  \int d^4\theta\frac{X^\dagger X}{M_{p\ell}^2}A^\dagger A,
  \hspace{3mm}
  \int d^4\theta\frac{X^\dagger +X}{M_{p\ell}}A^\dagger A,
 \end{eqnarray}
where $M_{p\ell}$ is the Planck mass. Then, the scale of soft terms
$F_X(\mbox{TeV}^2)/M_{p\ell}\sim
\mathcal{O}(10^{-3}$-$10^{-2}\mbox{eV})$ is expected. In the
``acceleressence'' scenario, this scale is identified with the dark
energy scale.\cite{acceleressence} We consider only one superfield 
which breaks supersymmetry for simplicity. If one extends the hidden 
sector, one can consider a different mediation mechanism between the 
standard model and the hidden sector from one between the dark and 
the hidden sector.

In this framework, taking supersymmetry breaking effect into account,
the scalar potential is given by
 \begin{eqnarray}
  V(\phi )=\frac{\lambda^2}{4}|\phi |^4-\frac{\kappa}{3}(\phi ^3+h.c.)
           +M_A^2|\phi |^2+m_D^2|\phi |^2-m^2|\phi |^2+V_0,
 \label{V1}
 \end{eqnarray}
where $\kappa$ and $m$ are supersymmetry breaking parameters, and $V_0$ is 
a constant determined by the condition that the cosmological constant
is vanishing at the true minimum of the acceleron potential.

We consider two types of the neutrino mass matrix in this scalar
potential. They are the cases of the vanishing and the
finite mixing between the left-handed and a sterile neutrino.

\subsection{Case of the Vanishing Mixing}

When the mixing between the left-handed and a sterile neutrino is
vanishing, $m_D=0$ in the neutrino mass matrix (\ref{MM}). Then we
have the masses of the left-handed and a sterile neutrino as
 \begin{eqnarray}
  m_{\nu _L} &=& c,\\
  m_\psi     &=& M_A+\lambda<\phi>. \label{sterilem}
\end{eqnarray}
In this case, we find that only the mass of a sterile neutrino is
variable on cosmological time scale due to the second term of the
right hand side in Eq. (\ref{sterilem}).

Let us adjust parameters which satisfy Eqs. (\ref{V}) and
(\ref{stationary2})$\sim$(\ref{amass}). In Eq. (\ref{V}), the scalar
potential Eq. (\ref{V1}) is used. Putting typical values for four
parameters by hand as follows:
 \begin{eqnarray}
  \lambda=1,\hspace{3mm}m_D=0,
  \hspace{3mm}m_{\nu _L}^0=2\times 10^{-2}\mbox{ eV},\hspace{3mm}
  m_\psi ^0=10^{-2}\mbox{ eV},
 \end{eqnarray}
we have
 \begin{eqnarray}
  &&<\phi>^0\simeq -1.31\times 10^{-5}\mbox{ eV},\hspace{3mm}
  c=2\times 10^{-2}\mbox{ eV},\hspace{3mm}
  M_A\simeq 10^{-2}\mbox{ eV},\nonumber\\
  &&m\simeq 10^{-2}\mbox{ eV},\hspace{3mm}
  \kappa\simeq 4.34\times 10^{-3}\mbox{ eV}.\label{value}
 \end{eqnarray}
We need fine-tuning between $M_A$ and $m$ in order to satisfy the
constraint on the present accerelon mass of Eq. (\ref{amass}).

\begin{figure}[ht]
\begin{center}
\includegraphics[width=0.8\linewidth]{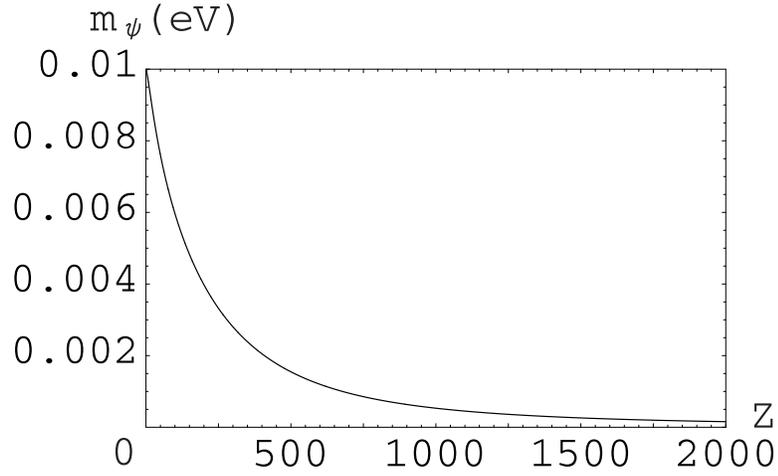}
\end{center}
\caption{Evolution of the mass of a sterile neutrino ($0\leq z\leq
2000$)}
\label{fig:2}
\end{figure}

\begin{figure}[ht]
\begin{center}
\includegraphics[width=0.8\linewidth]{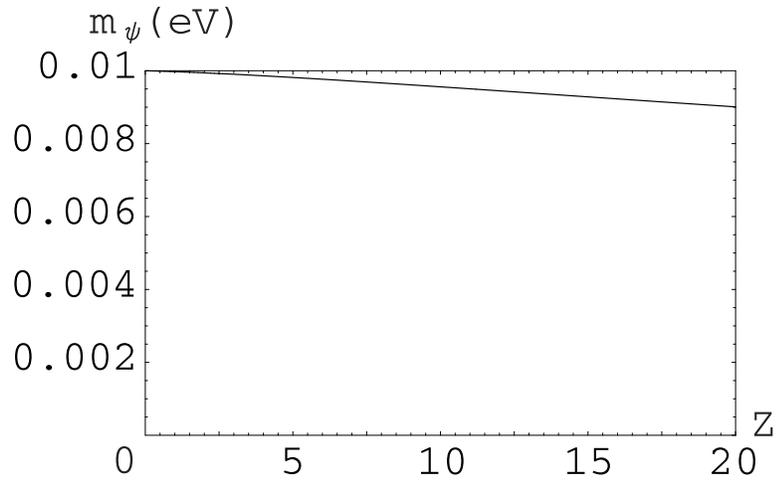}
\end{center}
\caption{Evolution of the mass of a sterile neutrino 
($0\leq z\leq 20$)}
\label{fig:3}
\end{figure}

\begin{figure}[ht]
\begin{center}
\includegraphics[width=0.8\linewidth]{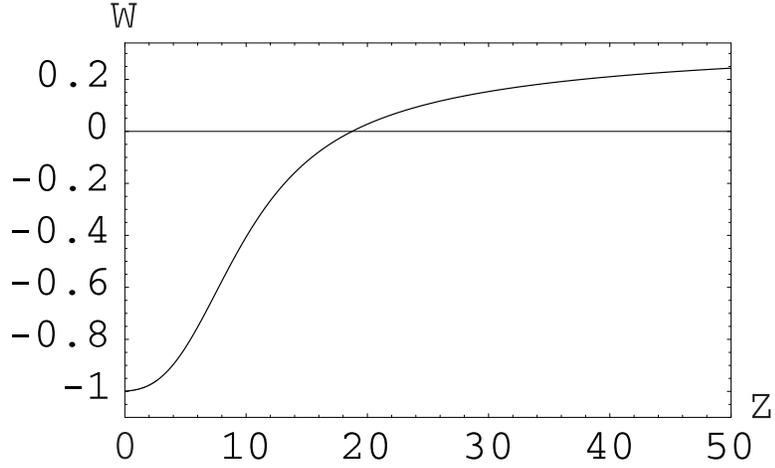}
\end{center}
\caption{Evolution of $w$ ($0\leq z\leq 50$)}
\label{fig:4}
\end{figure}

We show evolutions of the mass of a sterile neutrino and the equation 
of state parameter in Figs. \ref{fig:2}, \ref{fig:3} and \ref{fig:4}. 
The behavior of the mass of a neutrino near the present epoch is 
shown in Fig. \ref{fig:3}. We find that the mass  of a sterile 
neutrino have varied slowly in this epoch. This means that the first 
term of the left hand side in Eq. (\ref{constraint}), which is a 
negative contribution to the speed of sound squared, is tiny. We can 
also check the positive speed of sound squared in a numerical 
calculation. Therefore, the neutrino-acceleron fluid is 
hydrodynamically stable and acts as the dark energy.

\subsection{Case of the Finite Mixing}

Next, we consider the case of the finite mixing between the
left-handed and a sterile neutrino ($m_D\neq 0$). In this case, the
left-handed and a sterile neutrino mass are given by
 \begin{eqnarray}
  m_{\nu _L}&=&\frac{c+M_A+\lambda<\phi>}{2}
               +\frac{\sqrt{[c-(M_A+\lambda<\phi>)]^2+4m_D^2}}
                     {2},\\
  m_\psi&=&\frac{c+M_A+\lambda<\phi>}{2}
           -\frac{\sqrt{[c-(M_A+\lambda<\phi>)]^2+4m_D^2}}{2}.
 \end{eqnarray}
We find that both masses of the left-handed and a sterile
neutrino are variable on cosmological time scale due to the 
term of the acceleron dependence.

Taking typical values for four parameters as
 \begin{eqnarray}
  \lambda=1,\hspace{3mm}m_D=10^{-3}\mbox{ eV},\hspace{3mm}
  m_{\nu _L}^0=2\times 10^{-2}\mbox{ eV},\hspace{3mm}
    m_\psi ^0=10^{-2}\mbox{ eV},
 \end{eqnarray}
we have
 \begin{eqnarray}
  &&<\phi>^0\simeq -1.31\times 10^{-5}\mbox{ eV},\hspace{3mm}
    c\simeq 1.99\times 10^{-2}\mbox{ eV},\hspace{3mm}
  M_A\simeq 1.01\times 10^{-2}\mbox{ eV},\nonumber\\
  &&m\simeq 1.02\times 10^{-2}\mbox{ eV},\hspace{3mm}
  \kappa\simeq 4.34\times 10^{-3}\mbox{ eV}.\label{value1}
 \end{eqnarray}
where we required that the mixing between the active and a sterile
neutrino is tiny. In our model, the small present value of the
 acceleron is needed to satisfy the constraints on the scalar
 potential in Eqs. (\ref{V}) and (\ref{stationary2}).

\begin{figure}[ht]
\begin{center}
\includegraphics[width=0.8\linewidth]{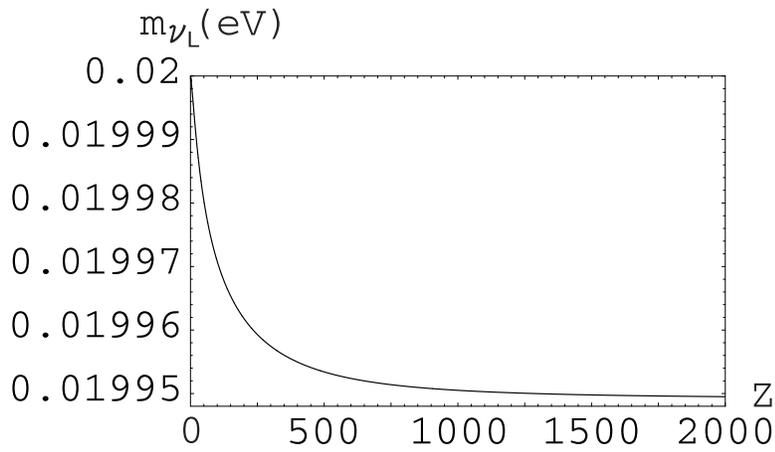}
\end{center}
\caption{Evolution of the mass of the left-handed neutrino
 ($0\leq z\leq 2000$)}
\label{fig:5}
\end{figure}

\begin{figure}[ht]
\begin{center}
\includegraphics[width=0.8\linewidth]{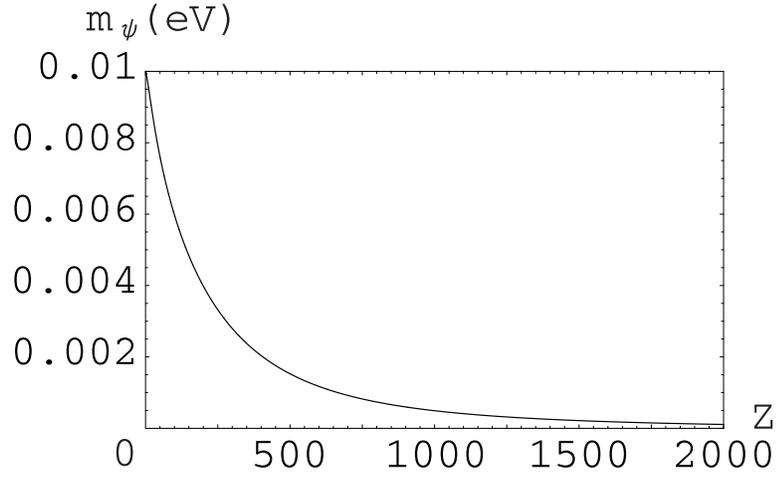}
\end{center}
\caption{Evolution of the mass of a sterile neutrino
 ($0\leq z\leq 2000$)}
\label{fig:6}
\end{figure}

\begin{figure}[ht]
\begin{center}
\includegraphics[width=0.8\linewidth]{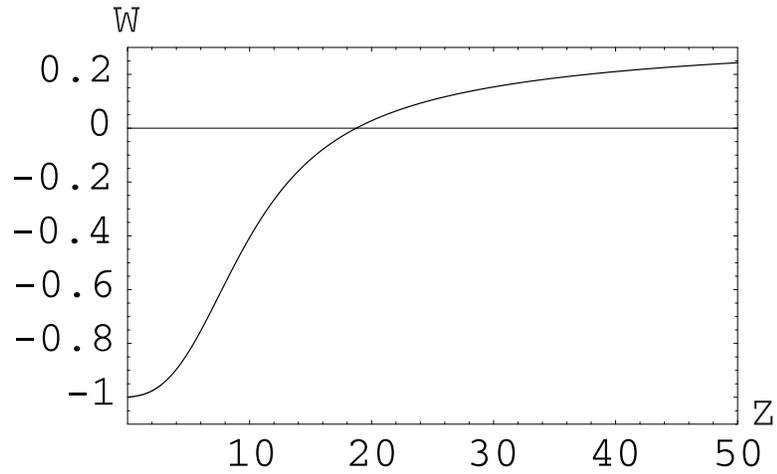}
\end{center}
\caption{Evolution of $w$ ($0\leq z\leq 50$)}
\label{fig:7}
\end{figure}

Values of parameters in (\ref{value1}) are almost same as the case of
the vanishing mixing (\ref{value}). However, the mass of the
left-handed neutrino is variable unlike the vanishing mixing case. The 
time evolution of the left-handed neutrino mass is shown in
Fig. \ref{fig:5}. The mixing does not affect the evolution of a
sterile neutrino mass and the equation of state parameter, which are
shown in Figs. \ref{fig:6}, \ref{fig:7}. Since the variation in the
mass of the left-handed neutrino is not vanishing but extremely small,
the model can also avoid the instability of speed of sound. 

Finally, we comment on the smallness of the evolution of the neutrino
mass at the present epoch. In our model, the mass of the left-handed
and a sterile neutrino include the constant part. A variable part is a
function of the acceleron. In the present epoch, the constant part
dominates the neutrino mass because the present value of the acceleron
should be small. This smallness of the value of the acceleron is
required from the cosmological observation and the stationary
condition in Eqs. (\ref{V}) and (\ref{stationary2}). 

\section{Summary}
We presented a supersymmetric MaVaNs model including effects of the 
supersymmetry breaking mediated by the gravity. Evolutions of the 
neutrino mass and the equation of state parameter have been calculated
in the model. Our model has a chiral superfield in the dark sector,
whose scalar component causes the present cosmic acceleration, and the
right-handed neutrino superfield. In our framework, supersymmetry is
broken in the hidden sector at TeV scale and the effect is assumed to
be transmitted to the dark sector only through the gravity. Then, the 
scale of soft parameters of
$\mathcal{O}(10^{-3}$-$10^{-2})(\mbox{eV})$ is expected. 

We considered two types of model. One is the case of the vanishing
mixing between the left-handed and a sterile neutrino. Another one is
the finite mixing case. In the case of the vanishing mixing, only the
mass of a sterile neutrino had varied on cosmological time scale. In
the epoch of $0\leq z \leq 20$, the sterile neutrino mass had varied
slowly. This means that the speed of sound squared in the neutrino
acceleron fluid is positive, and thus this fluid can act as the dark
energy. In the finite mixing case, the mass of the left-handed
neutrino had also varied. However, the variation is extremely small
and the effect of the mixing does not almost affect the evolution of
the sterile neutrino mass and the equation of state
parameter. Therefore, this model can also avoid the instability.



\end{document}